\documentclass[sigconf,screen]{acmart}
\usepackage{dirtytalk}
\usepackage{subfig}
\usepackage{multirow}

\AtBeginDocument{%
  }

\copyrightyear{2026}
\acmYear{2026}
\setcopyright{cc}
\setcctype{by}
\acmConference[CHI '26]{Proceedings of the 2026 CHI Conference on Human Factors in Computing Systems}{April 13--17, 2026}{Barcelona, Spain}
\acmBooktitle{Proceedings of the 2026 CHI Conference on Human Factors in Computing Systems (CHI '26), April 13--17, 2026, Barcelona, Spain}
\acmDOI{10.1145/3772318.3791970}
\acmISBN{979-8-4007-2278-3/2026/04}

\begin{document}
%
\title{Division of Labor and Collaboration Between Parents in Family Education}


\author{Ziyi Wang}
\authornote{Both authors contribute equally.}
\affiliation{%
\institution{Beijing University of Civil Engineering and Architecture}
  \city{Beijing}
  \country{China}}
\email{oliviawangnetwork11@gmail.com}
\orcid{0009-0008-8144-2543}

\author{Congrong Zhang}
\authornotemark[1] 
\affiliation{%
\institution{Electrical Engineering and Computer Science, University of Queensland}
  \city{Brisbane}
  \country{Australia}}
\email{congrong.zhang@uq.net.au}
\orcid{0009-0009-8076-2307}

\author{Jingying Deng}
\affiliation{%
  \institution{New York University}
  \city{New York}
  \state{NY}
  \country{USA}}
\email{jd5403@nyu.edu}
\orcid{0009-0001-0082-3491}

\author{Xiaofan Hu}
\affiliation{%
  \institution{Hangzhou Dianzi University}
  \city{Zhejiang}
  \country{China}}
\email{i@huxiaofan.com}
\orcid{0009-0005-1010-4510}

\author{Jie Cai}
\authornote{Corresponding author.}
\affiliation{%
\institution{Department of Computer Science and Technology, Tsinghua University}
  \city{Beijing}
  \country{China}}
\email{jie-cai@mail.tsinghua.edu.cn}
\orcid{0000-0002-0582-555X}

\author{Nan Gao}
\authornotemark[2] 
\affiliation{%
  \institution{Nankai University}
  \city{Tianjin}
  \country{China}}
\email{nan.gao@nankai.tsinghua.edu.cn}
\orcid{0000-0002-9694-2689}

\author{Chun Yu}
\affiliation{%
  \institution{Tsinghua University}
  \city{Beijing}
  \country{China}}
\email{chunyu@tsinghua.edu.cn}
\orcid{https://orcid.org/0000-0003-2591-7993}

\author{Haining Zhang}
\authornotemark[2] 
\affiliation{%
  \institution{Nankai University}
  \city{Tianjin}
  \country{China}}
\email{zhanghaining@nankai.edu.cn}
\orcid{0009-0007-1320-3981}






\renewcommand{\shortauthors}{Wang et al.}

\begin{abstract}
    Homework tutoring work is a demanding and often conflict-prone practice in family life, and parents often lack targeted support for managing its cognitive and emotional burdens. Through interviews with 18 parents of children in grades 1–3, we examine how homework-related labor is divided and coordinated between parents, and where AI might meaningfully intervene. We found three key insights: (1) Homework labor encompasses distinct dimensions: physical, cognitive, and emotional, with the latter two often remaining invisible. (2) We identified father-mother-child triadic dynamics in labor division, with children’s feedback as the primary factor shaping parental labor adjustments. (3) Building on prior HCI research, we propose an AI design that prioritizes relationship maintenance over task automation or broad labor mitigation. By employing labor as a lens that integrates care work, we explore the complexities of labor within family contexts, contributing to feminist and care-oriented HCI and to the development of context-sensitive coparenting practices. 

\end{abstract}

\ccsdesc[500]{Human-centered computing}
\ccsdesc{User studies}

\keywords{Collaboration, Children/Parents, Family Education, Qualitative Methods, Division of Labor, Human-AI Collaboration}

\maketitle

\section{Introduction}
Imagine this scene: a mother sits at the kitchen table late at night, trying once again to explain a homework problem to her child. Each repeated explanation tests her patience. What she is facing is not just solving a problem, but a layered mix of emotional, cognitive, and physical tension. Surveys show that about two-thirds of parents in \textit{Organization for Economic Co-operation and Development} countries are actively involved in their children’s homework~\cite{OECD2023PISA}. This seems like a routine part of parenting. Beneath this routine, however, lies a heavy load of psychological and emotional labor. Parents prepare materials and manage schedules, and also serve as emotional buffers, calming their child’s frustration while suppressing their own rising stress. Over time, a simple tutoring moment becomes a slow emotional struggle, turning homework tutoring into a \say{second shift} that stretches parents’ resilience~\cite{Hochschild2012ManagedHeart}.

Homework tutoring struggles are especially pronounced in China, where intense educational competition and heavy homework loads shape everyday family life. Surveys show that around 80\% of Chinese parents feel anxious about their children’s academic performance~\cite{MaZhang2024BlueBook}, and 67.5\% of mothers take the primary responsibility for homework tutoring~\cite{FourthSurveyWomenChina2021}. Weekly parental tutoring time for primary-school children has increased from 3.7 to 5.9 hours~\cite{FourthSurveyWomenChina2021}. This burden, however, is not distributed evenly. Gendered divisions of labor usually mean that mothers perform most of the cognitive and emotional work of homework tutoring, while fathers often take on peripheral roles involving occasional supervision~\cite{chen2024gendered}. These patterns not only exhaust families but also reinforce broader gender inequalities.

In light of these challenges, technological advancements, particularly in Artificial Intelligence (AI), present new opportunities. Prior work shows that AI can support emotional communication and provide educational assistance within families~\cite{kwon2022captivate, theofanopoulou2022purrble,chen2025scores, ho2025setpaired, viswanathan2025Interaction}. For instance, AI tools can ease parental emotional tension and facilitate intra-family communication~\cite{garg2020conversational,beneteau2020parenting}. 
Current systems, however, focus primarily on children’s learning needs~\cite{cagiltay2023homeworkbuddy,michaelis2018readingsocially, venugopalan2025caregiver} or parents'  attitudes toward technology~\cite{ho2025setpaired, petsolari2024sociotechnical, yu2025parentchildgai, garg2020conversational}.
They rarely address how labor in homework tutoring is divided or how gendered expectations shape everyday coordination. Therefore, current tools seldom engage with imbalances in caregiving responsibilities. Our study addresses this gap through two research questions (RQs):
\begin{itemize}
    \item \textit{RQ1.} How is the labor of homework involvement currently divided and coordinated between parents in Chinese families? 
    \item \textit{RQ2.} How can AI systems support more equitable parent collaboration, and what challenges and opportunities arise in this process?
\end{itemize}

To answer these questions, we conducted semi-structured interviews with 18 Chinese parents with children in grades 1–3. Building on feminist, care-oriented HCI work~\cite{bardzell2010feminist,daminger2019cognitive,roemmich2023emotion}, we use labor as an analytic lens to reveal that the division of labor operates as a dynamic, triadic cycle, where children's feedback triggers recurring cycles of breakdown and renegotiation. We propose shifting the design goal of AI from automating instructional tasks to fostering relational support. We outline a design agenda for family AI with three implications: collaborative annotation, child-centered narrative timelines, and joint reflection. These strategies aim to highlight the often invisible emotional and cognitive labor within families.

We extend research on family-centered HCI and Computer-Supported Cooperative Work (CSCW) in care by revealing the granular forms of labor and specific asymmetries in parental division in homework tutoring, and by offering actionable AI design implications that facilitate value alignment and promote sustainable coparenting arrangements.

\section{Background}
In HCI research, homework tutoring is often framed as a learning environment that requires safety and scaffolding. Prior studies have explored how technologies can help children cultivate self-regulation, ease parental supervision burdens, and foster positive parent-child interactions~\cite{neustaedter2006awareness,neustaedter2009calendar,luckin2016intelligence,holmes2019ai}. Yet much of this work is situated in Western contexts, where homework loads are relatively manageable, and parents often act as companions or supporters rather than tension managers~\cite{oecd2014pisa,PatallCooperRobinson2008HomeworkSynthesis}. In contrast, scholarship on East Asia, especially China, documents stronger parental emphasis on academic performance and high achievement expectations, which tend to produce more directive, management-like involvement in children's study routines~\cite{ng2007european,pomerantz2009role,chao1994beyond}. This cross-cultural difference suggests that design assumptions derived from Western settings may under-specify the needs of families who face heavier homework demands and performance-oriented parental oversight.

In China, homework tutoring has evolved into a high-pressure and enduring \say{second shift}~\cite{hochschild2012second}: dining tables quickly become desks as parents take on the roles of tutor, supervisor, and emotional mediator. Heavy workloads and broad subject coverage~\cite{oecd2014pisa}, compounded by children's volatile emotions, from distraction and procrastination to complaints and tears, transform this daily ritual into fragmented yet relentless labor. Crucially, this work is unevenly distributed: mothers typically take the core cognitive labor of planning, coordinating, and repairing emotional ruptures, while fathers' contributions are more occasional and substitutive~\cite{daminger2019cognitive,weeks2025typology,haupt2024gendered}. This structural imbalance leaves mothers chronically exhausted in time and energy, while silently accumulating tension within families~\cite{dean2022mentalload}. As such, in China, homework tutoring is not merely an educational practice but a mirror reflecting household power structures and the division of invisible work. For HCI, this context underscores a broader opportunity: to move beyond performance-oriented interventions and instead design for making hidden forms of labor visible, negotiable, and more evenly shared~\cite{star1999invisible}.

\section{Related Work}
\subsection{Labor and Coordination in Parenting}
Labor has long been a critical lens for understanding collaboration in HCI. Beyond visible and physical work, researchers have emphasized the emotional labor~\cite{Hochschild2012ManagedHeart}, invisible work~\cite{star1999invisible}, and cognitive labor~\cite{daminger2019cognitive} in coordination. These forms of labor, anticipating needs, planning trajectories, monitoring progress, and repairing breakdowns, rarely yield tangible outputs, yet they profoundly shape how responsibilities are allocated and how relationships are maintained~\cite{star1999invisible,daminger2019cognitive}. This lens has informed studies that render \say{invisible labor} more legible and negotiable through task visualization, collaboration scaffolds, and intelligent scheduling~\cite{neustaedter2006awareness,neustaedter2009calendar,bodker2013calendars,verdezoto2021maintenance,saxena2022unpacking}. However, this line of work has largely centered on formal, compensated settings~\cite{crain2016invisible}, leaving the informal, emotionally charged domain of family life comparatively underexplored.

Homework tutoring provides a concrete site for extending a labor lens in HCI to family learning contexts. It spans physical labor (e.g., preparing materials), emotional labor (e.g., soothing and motivating children), and cognitive labor (e.g., planning, coordinating, and monitoring tasks). Among these, cognitive labor is the least visible yet most consequential for structuring household divisions of responsibility and shaping parent-child relationships~\cite{daminger2019cognitive}. Prior work has documented significant gender asymmetries: mothers disproportionately take sustained \say{core} tasks, while fathers more often contribute sporadically with \say{episodic} involvement~\cite{weeks2025typology}. This division reflects deep power dynamics and societal expectations surrounding caregiving, underscoring the need for an equitable division of both emotional and cognitive labor at home~\cite{bardzell2010feminist}. While this pattern is evident across various cultural contexts, the high-frequency and emotionally intense nature of homework tutoring makes these dynamics both more visible and impactful~\cite{xu2005homework,pomerantz2005mothers,moe2018parents,li2024effects}. Yet HCI has paid limited attention to how this cognitive labor is reproduced, negotiated, or resisted in domestic learning environments, or to how technologies might redistribute rather than simply mask it.

At the same time, many AI-based educational technologies and intelligent tutoring systems deployed in homes have emphasized automating physical and evaluative labor, such as grading~\cite{luckin2016intelligence} or explanation~\cite{holmes2019ai}. While efficiency gains are valuable, 
these systems typically treat technology as a neutral instructional tool and overlook the cognitive and emotional dimensions that make homework a lived form of labor. This blind spot is particularly consequential in contexts such as China, where heavy homework loads and gendered cultural expectations converge to make mothers the de facto coordinators~\cite{hook2015making}. Here, homework is not merely an academic task but a site where invisible labor, gendered responsibilities~\cite{daminger2019cognitive}, and affective dynamics intersect~\cite{dean2022mentalload}. Addressing this requires HCI to move beyond efficiency-driven logics toward design mechanisms that help families identify, externalize, and rebalance hidden labor, positioning AI not as a tutor but as a scaffold for coordination, repair, and more equitable collaboration within families~\cite{toombs2018supporting}.

\subsection{AI-Supported Parent-Child Interaction}
AI-supported parent-child interaction has become a growing theme in HCI, with applications ranging from emotional companionship to educational support. Early studies have explored how smart speakers, chatbots, and social robots can support children's learning at home, but parents tend to approach these AI tools with more caution, emphasizing trustworthiness, safety, and the need for supervision~\cite{yu2025parentchildgai, garg2020conversational}. For example,  parents see potential benefits in AI assistants (such as objective feedback or help managing emotions), but worry that such technology could undermine their authority or encourage \say{lazy parenting} if it replaces parent-child interaction~\cite{petsolari2024sociotechnical}. Deploying Amazon Echo in family homes could both stimulate communication and trigger new tensions around control and fair access, highlighting the need for better support of parental mediation roles~\cite{beneteau2020parenting}. Similarly, social robots have been introduced into homework and shared reading contexts, and parents expected these robots to act as collaborators, not replacements, thereby deeply shaping parental expectations~\cite{ho2025setpaired}. While children valued robots' patience and encouragement, parents were more concerned with their ability to foster autonomy and sustained motivation in learning~\cite{cagiltay2024toward}.

Unlike playful or voluntary learning scenarios, homework tutoring often involves high levels of emotional tension and resistance from both children and parents, thus often filled with conflict~\cite{gao2025homework}. Recent systems have responded by augmenting not just cognitive but also emotional and social aspects of parent-child interaction, such as offering contextual language suggestions to support parent-child verbal interaction \cite{kwon2022captivate} and deploying the Purrble toy to prompt spontaneous emotional dialogue within families \cite{theofanopoulou2022purrble}. These efforts demonstrate AI's potential to support relational and affective dynamics during educational activities.

Through studies of parent-child interaction across different settings, most prior work has focused either on children's perspectives, parents' attitudes toward technology, or contrasting views between parents and children. Narrowing this lens to AI-assisted homework tutoring, we find similar research trends. In studies of social robots for homework support, researchers have primarily examined children's perceptions of such robots~\cite{michaelis2018readingsocially}. Venugopalan et al. extended this focus to the parental perspective, analyzing how caregivers interpret and respond to large language model (LLM) based systems that provide conversational support for homework assistance~\cite{venugopalan2025caregiver}. Likewise, Quan et al. investigated both children's and parents' views of ChatGPT, highlighting how their differing expectations shape the role of AI tools in family learning environments~\cite{quan2024chatgpt}.

In these studies, parents are typically treated as a single unit, overlooking the coordination and division of roles between caregivers. This dyadic framing obscures the uneven division of roles frequently observed in Chinese families, where one parent, most often the mother, assumes the majority of cognitive and emotional labor~\citep{daminger2019cognitive,weeks2025typology,haupt2024gendered}. While some HCI research has begun to interrogate the gendered dimensions of digital labor~\citep{roemmich2023emotion}, most AI-supported educational systems continue to treat parenting roles as interchangeable, thereby missing the opportunity to address persistent inequalities in caregiving responsibilities. To fill this gap, our study examines how AI can mediate not only parent-child interaction but also mother-father coordination, fostering balanced division of roles and effective family communication during homework tutoring.

\section{Methodology}
This study examines homework tutoring in the early primary years (Grades 1–3), a critical transition from a play-based kindergarten environment to formal schooling, when children begin developing independent learning habits and require substantial parental scaffolding. Prior research shows that parents’ homework tutoring during this stage is crucial for fostering positive academic and socio-emotional development~\cite{ElNokali2010} and that parents’ homework tutoring typically intensifies immediately after school entry~\cite{Tao2019}.

To capture the complexity of this formative period, we conducted in-depth semi-structured interviews to understand how parents divide and coordinate their responsibilities, as well as their expectations and concerns regarding the involvement of AI technologies. Positioned at the intersection of labor-division-AI design, our work extends the research traditions of cognitive labor~\cite{daminger2019cognitive}, invisible labor~\cite{star1999invisible}, and family-centered HCI~\cite{neustaedter2006awareness,neustaedter2009calendar}. We particularly examine how divisions of labor are reproduced through cycles of breakdown, substitution, and repair, and how technologies might serve as scaffolds at moments of collaborative rupture.

\subsection{Recruitment and Screening}
We designed and distributed a screening survey to identify suitable participants. Beyond serving as a sampling tool, the survey also advanced our research logic, transforming core concepts into operational dimensions. It focused on four areas: 
\begin{enumerate}
    \item \textbf{Family structure and roles}, identifying primary caregivers and accounting for specific cases such as single-parent households, thereby centering our analysis on the dynamics of division of labor. 
    \item \textbf{Division of labor and task types}, mapping concrete activities (e.g., supervising homework, explaining problems, managing time, providing emotional support, coordinating with partners) and their frequency to categories of cognitive, emotional, and physical labor, ensuring that our sample included families where at least one parent consistently bore tutoring responsibilities. 
    \item \textbf{Burden and tension}, assessing satisfaction levels and sources of stress (e.g., time conflicts, parent-child disputes, partner disagreements, school-family communication pressures) to identify real points of conflict, prioritizing families that had experienced such challenges. 
    \item \textbf{Technology use and adoption willingness}, investigating whether parents had used tools such as homework apps, learning devices, or voice assistants, and probing their visions of AI's potential role, which informed directions for follow-up questions in interviews. 
\end{enumerate}

\begin{table*}[t]
\centering
\begin{tabular}{llll}

\toprule
\textbf{ID} & \textbf{Gender} & \textbf{AI Products in Use} & \textbf{Main Supervisor} \\
\midrule
P1  & mother & photo-based homework-solving app, educational tablet, AI voice assistant, AI chatbot & mother \\
P2  & mother & photo-based homework-solving app, AI voice assistant, AI chatbot & mother \\
P3  & mother & photo-based homework-solving app, AI chatbot & mother \\
P4  & father & photo-based homework-solving app, AI voice assistant & father \\
P5  & father & photo-based homework-solving app, educational tablet, AI chatbot & mother, father \\
P6  & mother & photo-based homework-solving app, AI voice assistant & mother \\
P7  & mother & photo-based homework-solving app, AI voice assistant, AI chatbot & mother \\
P8  & mother & photo-based homework-solving app, AI chatbot & mother \\
P9  & mother & photo-based homework-solving app, educational tablet, AI voice assistant & father \\
P10 & father & photo-based homework-solving app, educational tablet, AI voice assistant & mother, father \\
P11 & father & photo-based homework-solving app, AI voice assistant, AI chatbot & mother \\
P12 & mother & photo-based homework-solving app, AI chatbot & mother \\
P13 & father & photo-based homework-solving app, AI voice assistant & mother \\
P14 & mother & educational tablet, AI voice assistant & mother, father \\
P15 & mother & photo-based homework-solving app, AI voice assistant, AI chatbot & mother \\
P16 & mother & photo-based homework-solving app, educational tablet, AI voice assistant & mother, father \\
P17 & father & photo-based homework-solving app, AI voice assistant, AI chatbot & mother, father \\
P18 & mother & AI chatbot & father \\
\bottomrule
\end{tabular}
\caption{Parent participants, AI products in use, and main homework supervisors.\protect\footnotemark}
\label{tab:participants_parents}
\end{table*}

\footnotetext{
    Participants referred to specific Chinese AI tools: 
    (1) Photo-based apps: \textit{Zuoyebang} (\url{https://www.zuoyebang.com}) and \textit{Xiaoyuan Souti} (\url{https://www.yuansouti.com/}), which use OCR to solve problems similar to \textit{Photomath}; 
    (2) Educational hardware: \textit{iFlytek} tablets and \textit{Youdao} dictionary pens; 
    (3) Voice assistants: \textit{Xiao Ai} (Xiaomi) and \textit{Tmall Genie} (Alibaba); 
    (4) AI chatbots: \textit{Doubao} (\url{https://www.doubao.com/chat/}) and \textit{DeepSeek} (\url{https://www.deepseek.com/en/}), which are LLMs similar to \textit{ChatGPT}.
}

Based on this survey, we distributed recruitment messages via parent WeChat groups and social media platforms (e.g., Xiaohongshu) and screened respondents against our criteria. In total, 133 questionnaires were collected. The first author initially filtered the valid responses based on the completion time (over 90 seconds), parents’ moderate-to-high frequency of homework tutoring, challenges encountered during tutoring, and prior experience using AI products. After contacting eligible parents to arrange interview times, 18 parents were ultimately selected and completed the interviews, including 12 mothers and 6 fathers (see Table~\ref{tab:participants_parents}).

For the main supervisors involved in homework tutoring, 10 households were mother-only, 5 involved both parents, and 3 were father-only. This participation pattern highlights the predominant role of mothers in homework tutoring, followed by joint participation, with father-only involvement being comparatively rare.

\subsection{Semi-Structured Interview}
Guided by our RQs, the semi-structured interview protocol was organized into six themes: (1) current practices of homework tutoring and division of roles, (2) digital labor and \say{invisible work}, (3) emotional labor and psychological experiences, (4) partner involvement and family structures, (5) experiences with existing AI/technological tools, and (6) trust, potential, and fairness in AI design. The first three themes primarily addressed RQ1, while the latter three focused on RQ2.

To ensure a shared understanding of AI amidst varying levels of technical literacy, we framed the concept by referencing the specific algorithmic tools participants reported using in the screening stage (e.g., Zuoyebang, ChatGPT, Xiaoyuan). We defined AI in this context as technologies that intervene in the tutoring process through recognition, voice interaction, or problem-solving, anchoring their responses in concrete user experiences rather than abstract definitions.

The 18 participants were divided into three groups (7, 6, and 5), with each group interviewed by one of the first three authors. Considering the potential power dynamics and biases involved in the division of labor, all interviews were conducted in a one-on-one format. For the initial five interviews, at least two interviewers were present to calibrate and align interviewing styles. Each interview lasted between 40 and 90 minutes. With participants' consent, all interviews were audio-recorded. Participants were compensated at a rate of 100 RMB/hour. This study was reviewed and approved by the institutional ethics committee. 

Midway through the interview process, the research team held meetings to refine the interview guide, aiming to explore the nuances of labor division more deeply and uncover participants’ emotions. For instance, we adjusted the question sequence to encourage participants to open up progressively, moving from general to more in-depth discussions. Broad questions such as \say{\textit{What topics do you and your partner usually discuss when tutoring your child?}} were expanded into more detailed sub-questions like \say{\textit{What kinds of conversations occur between the father and the child, the mother and the child, and between the parents themselves?}}. All interviews were conducted and analyzed in Mandarin. Only selected quotes were translated for reporting to ensure that the data preserved their original meanings as accurately as possible.

We followed Francis et al.’s data saturation principle~\cite{francis2010saturation}, which defines saturation as the point when no new themes emerge across three consecutive interviews after an initial set of ten. In our study, thematic saturation was reached after analyzing the eighteenth parent interview. After all interviews were completed, the research team met again to organize, cross-check, and share findings from each interview.

\subsection{Analytical Approach}
The interview data were analyzed collaboratively by the first three authors. We adopted a thematic analysis approach with inductive coding~\cite{patton1990qualitative}. All interviews were audio-recorded, transcribed, and analyzed using Braun and Clarke's six-phase method of thematic analysis~\cite{braun2006using}. 

We adopted an inductive coding approach involving three phases. In the initial phase, to ensure inter-rater reliability, the first three authors collectively coded one interview transcript to establish a unified initial codebook. Subsequently, three authors independently analyzed the remaining interviews conducted by the other two, allowing cross-checking and reducing self-bias. We convened weekly to discuss new codes and resolve discrepancies. In the consolidation phase, all preliminary codes were aggregated on a shared Miro board. Through affinity diagramming, we iteratively clustered codes into subcategories and synthesized them into broader themes.

Finally, in the refinement phase, we aligned these themes with our RQs. Emerging topics that did not directly address the RQs were critically reviewed and excluded to maintain focus. For instance, while many parents discussed the burden of non-academic assignments required by schools (e.g., crafting and hand-copying newspapers) and the resulting home-school tensions, these codes were excluded because they did not pertain to the internal dynamics of parental division of labor or coordination. This process crystallized into two high-level categories: (1) parental division of labor and (2) perceptions and expectations of AI, which encompass six specific subthemes detailed below.

This collaborative, iterative process, combining independent coding, joint clustering, and cross-validation, strengthened the credibility and rigor of the analysis. Triangulation across coders enhanced analytical trustworthiness, while the joint synthesis process allowed the team to move beyond surface-level codes and uncover deeper conceptual patterns. Importantly, the approach enabled us to develop a situated understanding of how parents conceptualize the division of labor and imagine the potential role of AI in reshaping these family dynamics.

\section{Findings}
\subsection{The Construction of Parental Division of Labor: Structural, Interpersonal, and Child-Driven Dynamics}
\label{sec:5.1}
To address RQ1, we examined how parents negotiate the division of labor in homework tutoring. This division is rarely a result of free individual choice; instead, it is a dynamic outcome shaped by three distinct forces: external structural constraints (economy and work), gendered expectations (gendered perceptions of patience and expertise), and the agency of the child.

\subsubsection{Structural Constraints Regarding Economy and Work}
\label{sec:5.1.1}
A dominant pattern observed across the participating families was the shaping of labor division by external institutional pressures. Work-related demands, such as overtime, distant postings, and frequent business travel, structurally limited parents' availability. Furthermore, school-assigned weekend tasks, digital platforms for home-school communication, and rigid extracurricular training schedules introduced additional layers of \say{digital labor} (Mother, P2 and P16; Father, P11). Collectively, these systemic conditions positioned mothers as the primary tutors.

Participants frequently rationalized this imbalance through an economic lens. In several cases, the lower-earning parent, typically the mother, was expected to reduce professional involvement and assume daily caregiving. As one mother reflected, echoing a sentiment shared by participants in dual-income households, this logic was grounded in the notion of \say{cost-effectiveness}: \say{\textit{With the same amount of time, he can create more value than I can. From the family’s perspective, my contribution isn’t as high as his, so he thinks I should work less and spend more time with the children.}} (Mother, P8). This economic rationalization often led mothers to internalize the role of primary caregiver. Mother P3 is a full-time homemaker who takes care of all household chores and matters related to her child. She explained that she had gradually internalized this division, even expressing empathy toward her husband's workload: 

\begin{quote}
\say{\textit{I’ve just gotten used to this setup. I know he works really hard, so I don’t want him to have to worry about things at home or the child’s education as well. At least he can support us, so I can stay home and take care of our child without having to work a full-time job. I still think this arrangement makes sense.}} (Mother, P3)
\end{quote}

Apart from economic reasoning, the physical constraints of work schedules, such as overtime and business travel, often rendered \say{fair} division impossible. When fathers were absent due to long-term working in different locations, mothers reported taking on the entire responsibility for homework tutoring, including ongoing supervision. P14 noted the structural constraint of her husband's prolonged absence:

\begin{quote}
\say{\textit{My husband works for the state grid, so he travels a lot for work... Sometimes he’s gone for three or four months, and I’m the only one looking after our child.}} (Mother, P14)
\end{quote}

This constraint disproportionately impacted mothers' professional aspirations. Several mothers described how household tasks and tutoring drained their energy, undermining their ability to pursue professional development. P6 reflected, \say{\textit{If I didn't have children or a family, I’d probably have the energy to compete and chase more opportunities. But now so much of my energy goes into childcare and family that my career is kind of held back.}} (Mother, P6). 
P6 offered a vivid account of the exhaustion caused by balancing professional and domestic roles: 

\begin{quote}
\say{\textit{I’m a teacher, and I’m basically working all the time. I take care of the child, do my job, and even after work, I still have to soothe the child, do the housework, and then prepare for my public classes. A lot of nights I’m up until 4 a.m. Sometimes I only get two hours of sleep. When I’m teaching, I honestly feel like I’m about to collapse once the class is over. There was news about a teacher fainting on stage, and I really think that’s possible, because teachers are just too exhausted.}} (Mother, P6)
\end{quote}
\subsubsection{Gendered Expectations Regarding Patience and Expertise}
\label{sec:5.1.2}
Beyond structural constraints, the division of labor was heavily influenced by interpersonal dynamics, specifically the negotiation of emotional labor and subject expertise.

\paragraph{Gendered Emotional Labor and ``Patience''}

While some participants attributed labor division to innate \say{personality}, our analysis suggests these attributions reflect broader gendered expectations of emotional labor.

Across several families, fathers were often described as more \say{impatient} or \say{quick-tempered} (Mother, P1, P6, P8; Father, P10, P11), which made them less suitable for the sustained patience required in tutoring. For instance, P1 recalled that the father's emotional outbursts undermined his ability to serve as the tutor:

\begin{quote}
\say{\textit{Yes, he’s pretty impulsive. We take turns helping our child with homework, but whenever it’s his turn, he ends up yelling things like, \say{Why can't you learn it?} and it makes our child cry a lot.}} (Mother, P1)
\end{quote}

P8 emphasized that her husband's temperament was fundamentally incompatible with tutoring tasks: \say{\textit{He’s quite impatient, so he doesn’t feel comfortable with detailed academic guidance, and he loses his temper with our child pretty easily.}} (Mother, P8).

Conversely, mothers frequently reported heightened anxiety about their children's academic performance, which drove them to intervene more intensively. For instance, P12 commented: 

\begin{quote}
\say{\textit{I feel okay about it, but his mom is a lot more anxious and thinks other children are doing better. She’s always talking with other moms about how competitive the class is and how many practice papers other children can finish in a day. After hearing all that, she gets influenced and comes back asking me to give our child more practice papers to do.}} (Father, P12)
\end{quote}

This dynamic was often exacerbated by conflicting educational philosophies. Fathers often emphasized children's innate ability and long-term development, while mothers focused on immediate academic outcomes. P10 explained, \say{\textit{Ability also matters, and I’ve always believed that if you improve your ability, the results will naturally follow.}} (Father, P10). In contrast, his wife described, 

\begin{quote}
\say{\textit{Yes, I’m really anxious. I often ask him what we should do since our child isn’t doing well at school, how we should adjust, and how we can help our child improve. But he’s not very interested in any of this. He just thinks that if the grades are bad, we’ll figure something out later. He even believes that talent is innate, that maybe our child just doesn’t have it, and that there’s not much parents can do to change that.}} (Mother, P3)
\end{quote}

These findings highlight how the cultural norm of \say{men outside, women inside} continues to shape the division of emotional labor. P3 remarked, \say{\textit{My husband doesn’t really pay much attention to our child’s studies. He mostly just asks about grades and how the child is doing. He rarely helps with schoolwork unless I’m very busy or not around, and then he’ll take over. Otherwise, he’s basically not involved.}} (Mother, P3). Another recalled that her husband framed her heavier involvement as \say{normal} simply because of gender: \say{\textit{He thinks that because I’m a woman, I should naturally do more at home, and that’s just how it is.}} (Mother, P7).

Rather than treating these descriptions as evidence of inherent personality differences, we interpret them as narratives that draw on gendered cultural norms in which fathers are permitted to step back from emotionally intensive work, while mothers are expected to absorb frustration, maintain patience, and protect the child's emotional well-being. In this sense, \say{patience} operates less as a neutral personality trait and more as a moralized expectation placed on mothers. These gendered expectations explain why mothers in our study carried the bulk of instructional work and also undertook the ongoing emotional regulation that sustained homework tutoring over time.




\paragraph{Subject-Based Expertise}
Subject expertise, unlike the emotionally charged negotiation of patience, was viewed as pragmatic. A recurring pattern assigned fathers responsibility for mathematics and science, while mothers took charge of Chinese and English. This specialization facilitated efficiency but also created fixed boundaries that rarely shifted once established (P4, P9, P10).

P9 described how she and her husband organized tutoring according to their expertise: 

\begin{quote}
\say{\textit{Sometimes his dad tutors him in math, which I feel is more rigid and straightforward, like one plus one equals two. I usually handle Chinese, which is more flexible and needs more explanation. So we’ve just split it that way: he does math, I do Chinese.}} (Mother, P9)
\end{quote}

P4 highlighted how this clear division was negotiated to reduce cognitive burden and distribute difficult tasks more evenly: \say{\textit{Since I already help him with math, I really want someone to share this part with me, especially the calculations.}} (Father, P4). 

Once established, these subject-based boundaries provided a relatively stable arrangement regarding labor division: \say{\textit{We split the work, with him doing math and me doing Chinese. Math is more straightforward; it’s either right or wrong. Language needs more guidance and interpretation, and that suits me better.}} (Mother, P10).


\subsubsection{The Child as an Active Agent of Division}
\label{sec:5.1.3}
Children are not passive recipients of care but active agents who reshape the division of labor. Children frequently acted as the \say{tie-breaker}, gravitating toward the parent with a clearer teaching style or better emotional regulation.

In multiple instances, a child's response directly prompted parents to renegotiate responsibilities. P4 explained that tutoring roles shifted pragmatically depending on the child's comprehension: \say{\textit{If he really can't understand when I explain, I just switch and let his mother teach him.}} (Father, P4). Similarly, children sometimes explicitly resisted a parent's guidance because of the parent's communication style. As P8 noted, \say{\textit{My son doesn’t like asking his dad for help. The way his dad explains things is hard for him to understand, sometimes confusing, and often not very step-by-step. He doesn’t really get it, so his dad ends up just giving instructions instead of actually teaching.}} (Mother, P8).

Furthermore, children’s dissatisfaction with the primary caregiver often exacerbates parental conflict. As Father P13 noted, children's dissatisfaction with the primary caregiver's teaching methods could intensify the parent-child relationship. When children rejected the mother's explanations, delivered amidst her household chores, it forced families to reconsider their division of labor (Mother, P15). Thus, children's voices and reactions acted as a feedback loop, prompting parents to renegotiate boundaries and adjust arrangements.

In Father P13’s family, both he and his partner graduated from a rural high school and did not feel equipped to tutor their child. He worked in a factory, so most of the time his partner, a homemaker, assumed this responsibility. As he explained, the child frequently resisted the mother's explanations, which were delivered while she was also burdened with household responsibilities: 

\begin{quote}
\say{\textit{Most of the time, the child feels dissatisfied. I’m busy with work, and she’s now a full-time mom handling daily chores and all kinds of small things. When the child has questions, she explains through the tablet, but he usually can’t understand, so he often ends up feeling frustrated.}} (Father, P13)
\end{quote}

In this account, dissatisfaction was not merely about the clarity of instruction but was intertwined with broader family arrangements. Since the mother was the primary caregiver, her tutoring was unavoidable and, from the child's perspective, at times less effective, leading to frustration for both sides.

Such dissatisfaction frequently escalated beyond the parent-child dyad, intensifying the couple's coordination. A brief remark from another participant captured this dynamic: \say{\textit{It’s mainly the child who’s unhappy. The child isn’t satisfied with how we divide the work.}} (Mother, P15). Here, the child's frustration with the quality of tutoring fed back into marital negotiations, showing how children's voices and preferences could renegotiate the division of labor. 

\subsection{Bridging Experience and Expectation: The Role of AI in Parenting}

To address RQ2, we analyze how parents’ current technology practices shape their expectations for future AI tools. Parents do not seek \say{automation} of parenting but rather \say{augmentation} that helps them manage the challenges identified in RQ1. We identify three core dimensions: (1) family coordination, (2) tutoring process, and (3) emotional regulation.

\subsubsection{From Invisible Labor to Data-Driven Coordination}
Parents currently lack tools to visualize the \say{invisible labor} of educational management. As noted in Section \ref{sec:5.1}, the burden of coordination often falls disproportionately on mothers. In response, parents consistently envisioned a role for AI that extends beyond direct pedagogy: acting as a mediator of caregiving. 

Families described how an effective system would \say{surface who should take over}, making implicit labor explicit. Mother P8, who works in K-12 educational product and content development and has a strong interest in educational product design, articulated three concrete design requirements. First, \emph{legibility}: summaries that quantify time, tasks, and cognitive load so that contributions become visible, enabling partners to both recognize effort and identify areas for optimization. Second, \emph{neutrality}: an even, data-driven tone that reframes \say{you should} as \say{here's what the data suggests}, positioning AI as a bridge rather than a judge. Third, \emph{timing}: situationally aware handoff hints and conflict-time guidance calibrated to social roles and temperament, with advice framed in a balanced manner.

\begin{quote}
\say{\textit{If there were some clear data, I think it would help in a few ways. First, it would let my partner really see what I’ve been doing. Second, it would help me see where I could do better. And third, it would give me a sense of achievement, or at least a visible record of what I contribute to the family. When all that effort is visible, it shows where things could be improved... and it also helps my husband see which parts could be shared or taken on by him.}} (Mother, P8)
\end{quote}

Parents envisioned AI as a socio-emotional intermediary that could detect imbalance and prompt the other partner to provide support. When envisioning these systems, participants did not describe embodied robots or simple chatbots. Instead, they conceptualized AI as a pervasive \say{ambient auditor}, a background system capable of aggregating data across devices to render an objective judgment on family contributions. P8 further advocated paired self- and partner-views that render contributions legible, identify plausible opportunities for complementary involvement, and attach handoff suggestions in neutral and ready-to-use terms. She emphasized that the guidance should be grounded in observable traces and in a facilitative manner, for example, who performed which tasks, with what frequency and duration, and the previous interaction history.

\begin{quote}
\say{\textit{One really good thing would be if AI could tell fathers, tell men, \say{Your wife has put in a lot, you should give her more care, and maybe take on more yourself.} It could act like a bridge between the two of us. It would need to understand what role each parent usually plays, whether they have time in the evenings, what their mood is like, and even how they want to split responsibilities. Then it could give suggestions based on what’s actually going on. For example, my work is really busy, and my husband has his own expectations. He wants me to spend more time, and I also want him to spend more time. When those expectations clash, what do we do? I think AI should really take into account parents’ roles in society, their personalities, and even their past experiences, and then give balanced, thoughtful advice.}} (Mother, P8)
\end{quote}

This mother suggested that AI could make visible the contributions of the partner, such as a mother's disproportionate investment in tutoring, and prompt the other partner to provide care and support. In this sense, AI is imagined as a \say{bridge} that facilitates communication and alleviates gendered tensions within the family. This expectation suggests that AI is valued not merely for tutoring the child, but for facilitating more equitable family responsibilities.

\subsubsection{From ``Answer-Getting'' to Scaffolded Learning}

Building on that primary expectation, they also described a complementary set of supports within homework tutoring itself. In this supplementary role, AI is imagined as an adult assistant that helps plan, scaffold, and, when desired, temporarily assume tutoring tasks while respecting parental authority. Rather than isolated features, parents portrayed a continuum of assistance, from homework planning for children to emotional regulation and takeovers of parents, each still emphasizing reasoning before answers, a transparent process, and close alignment with curricular materials.

\paragraph{Current Experience: Efficiency vs. Entertainment}
Parents identified advantages of current tools for improving efficiency and settling knowledge disputes. P3 emphasized that video-based explanations were easy for children to follow, so she bought the device:

\begin{quote}
\say{\textit{I think the AI video explanations are great, and my child can actually understand them. We’ve used them before, and this year I bought him a learning device. These new tools really reduce parents’ pressure and anxiety. As a parent, when I have tools to help me, I feel much less stressed.}} (Mother, P3)
\end{quote}

Tools functioned as complementary resources where parents perceived skill gaps. Parents highlighted the reading pen devices, allowing children to read along when adults cannot mimic the sound:

\begin{quote}
\say{\textit{What I find really useful is the reading pen. It works really well. For example, when my child doesn’t know an English word, they can just tap it and the pen reads it out loud, and the child can follow along. I think it’s actually great, because as parents, we don’t always know the correct pronunciation either.}} (Mother, P9)
\end{quote}

Tools that combine explanations with progress analytics help parents identify areas for improvement and deliver more targeted instruction. Parents reported that explanations were brief and usable, \say{\textit{a quick read and I can do the problems}} (Father, P5), and that the system could show key points and diagnose weak spots, while lightweight animations sustained interest (Mother, P1). P1 also noted higher voluntary study time when interactive elements and short games were present, with sessions stretching from 30 minutes to about an hour on weekends because the child \say{\textit{really liked it.}}

Generative drafting reduced effort during time-pressed moments by supplying exemplars that parents lightly edited before use. P7 described entering a prompt to obtain a model composition and then making small changes before asking the child to copy it when she was too busy or tired to write from scratch:

\begin{quote}
\say{\textit{When I’m busy, or when I’m already really tired and don’t feel like writing, I just use this app. I type in a prompt, let it generate an essay, then I make a few quick edits and give it to my child to copy.}} (Mother, P7)
\end{quote}

These reports converge on a practical profile of benefits: tools speed up routine explanations, expand child engagement, and provide caregivers with actionable visibility into areas where content is inaccurate or not curriculum-related. These affordances reduced parents' instructional load and made homework tutoring smoother.

However, parents expressed deep concern that some tools trade sustained learning for attention, pulling children toward entertainment rather than knowledge processing:
\begin{quote}
\say{\textit{One really good thing about Doubao is that it’s very engaging. It explains things with animated characters, so it feels fun and exciting. But the problem is that the child can get too absorbed in the animation and not really focus on the learning. Our child used it for about three months, and I felt the results weren’t great, definitely not as good as using a traditional learning device. After that, I felt it wasn’t very effective. The learning device feels more professional and more \say{nutritious,} while Doubao is more for entertainment. Most of the time, our child actually just uses Doubao to have fun, like telling jokes or for companionship.}} (Mother, P1)

\end{quote}

Parents also faced governance challenges with the tools: \say{\textit{When I am exhausted, taking it back from the child is actually a pretty tough process… the child isn't really willing to hand it over.}} (Mother, P6). Furthermore, generic answers not aligned with school requirements created extra \say{debugging} work for parents (Mother, P2).




\paragraph{Expectation: Process-Oriented Scaffolding} Parents demanded AI tools that prioritize reasoning over answers and ensure curricular alignment. Parents highlighted stable retrieval and step-by-step guidance as priority capabilities:
\begin{quote}
\say{\textit{First, the search feature needs to be reliable and cover enough content, so it can quickly find the exact question the child is looking for. Second, it shouldn’t just give the final answer. It should show the steps, and maybe even include a video explanation. Third, for each concept, it should branch into related problems so the child can practice and learn how to apply it.}} (Mother, P4)
\end{quote}

Checking homework was also a key efficiency tool: \say{\textit{Checking homework is helpful too. If there’s a tool where you can just scan it once and all the answers pop up, the whole process is really smooth.}} (Father, P10).

Tools should be anchored in textbooks and vetted item banks with an explainable rationale, provide parent controls that curb entertainment spillover, e.g., quotas, cooldowns, usage dashboards, and avoid over-anthropomorphism that might distort family roles; under these conditions, such tools function as scaffolds rather than substitutes, supporting learning without hijacking attention and assisting parent–child interactions without blurring role boundaries or displacing parental authority.

Parents emphasized procedural scaffolding that turns adult help into a routine. They asked for step-by-step guidance for caregivers on planning the evening's work, pacing attention and breaks, and turning momentary breakdowns into information for tomorrow's plan. Beyond individual assignments, families sought support that scales to routine design, which captures when and how accompaniment occurs, identifies what helps or hinders growth, and proposes concrete adjustments for the next session.

Parents described two core functions. First, planning aids tasks into sequenced, manageable chunks for the caregiver to supervise, e.g., \say{what to do first}, \say{when to pause}, \say{how to resume}, so the process feels guided rather than improvised (Mother, P6). Second, session logging records time use and pain points during accompaniment, enabling simple post-hoc reviews of \say{where it flowed} and \say{where it stuck} to inform the next day's plan, like P8 said:

\begin{quote}
\say{\textit{It could also track how I spend my time with my child. For example, every time I’m with them, it could record what was the hardest moment or what felt the most rewarding, and save those points.}} (Mother, P8)
\end{quote}


These parents wanted the product included in a plan the night before, with paced steps and scheduled breaks; a log of time and friction points; and the conversion of those traces into next-day adjustments that target the difficulty. The objective is for AI tools to coach adults to structure, observe, and iterate. 

A subset of parents explicitly asked for a \say{Full Handoff} option to relieve exhaustion, but even this was framed as a need for a reliable delegate. As one father emphasized, \say{\textit{Because once it takes over, I can go do my own thing.}} (P4), while another mother noted, \say{\textit{Anyone helping me share it would be fine. Because I'm always here, and I have to be patient, really patient.}} (P15).
\subsubsection{From Emotional Escalation to Regulated Intervention} The high emotional cost of tutoring drove cautious but specific expectations for emotional support. Parents acknowledged that fatigue often leads to rapid escalation of conflict, yet they were skeptical of AI replacing human connection. Some parents found human-like reassurance potentially helpful (Mother, P1 and P3), but others noted that during high tension, even human intervention fails (Mother, P7). P7 exemplifies this boundary case:

\begin{quote}
\say{\textit{When I’m about to lose my temper, honestly, even if a real person tried to calm me down, I probably wouldn’t listen. So you can imagine how much harder it would be for me to listen to an AI.}} (Mother, P7)
\end{quote}

Consequently, parents preferred lightweight, precisely timed prompts that respect adult agency. In this context, participants interpreted the \say{intelligence} of AI primarily as \say{emotional durability}. They defined the AI by its incapacity for anger, viewing it as a \say{cool-headed partner} that possesses the infinite patience biologically impossible for exhausted parents. The system would act as a \say{co-pilot} for regulation rather than a substitute parent. This necessitates human-in-the-loop oversight to ensure that AI interventions, such as reminders to stay calm, are delivered when they can be effectively received, without blurring role boundaries or displacing parental authority.

\section{Discussion}
We interpret our findings through a labor lens in the family context, aiming to surface structural dilemmas in homework-related collaboration rather than provoke gender antagonism or assign blame. We first clarified how physical, emotional, and cognitive labor operate in a triadic father-mother-child relationship, arguing for a shift from simply \say{making labor visible} to cultivating child-centered mutual understanding across recurring cycles of breakdown, repair, and renegotiation. Then we translate this lens into three design directions for AI in the family.

\subsection{Constructing the Labor Lens: The Hidden and Dynamic Nature of Parental Collaboration}

In our study, we focus on how parents divide and coordinate their work in home tutoring and characterize this labor along three distinct dimensions: physical, emotional, and cognitive labor. Physical Labor encompasses the instrumental tasks of execution, such as checking homework, grading exercises, compiling wrong-question sets, and preparing tutoring materials, and describes how these concrete tasks are allocated between parents. Emotional Labor centers on affective regulation and atmosphere maintenance, including soothing the child’s anxiety, providing encouragement, regulating parental anger, and sustaining a positive learning environment, often with one parent informally taking on a larger share of this burden. Cognitive Labor pertains to the managerial and organizational aspects of tutoring. It involves orchestrating the workflow, allocating parental responsibilities, coordinating schedules, monitoring academic progress, and identifying issues for strategic adjustment, and is key to understanding how roles and responsibilities are negotiated between mothers and fathers.

Among these three dimensions, physical labor has been the primary focus of existing commercial AI solutions, which excel at instrumental tasks such as automated grading and answer retrieval. Crucially, these tasks rarely necessitate complex mechanisms for the division of labor. Conversely, emotional and cognitive labor present unsolved challenges~\cite{LinEtAl2025Parental}. Although attempts have been made to address emotional labor via empathetic dialogue agents~\cite{Rashkin2018IKnowTheFeeling, Smith2020BlendSkills} and emotion detection~\cite{Mollahosseini2019AffectNet, Busso2008IEMOCAP}, the quality of such support remains a subject of debate~\cite{AINow2019Report}. Moreover, cognitive labor is rarely discussed in the context of AI design, largely because it requires navigating intricate negotiations and decision-making processes between parents and children. Therefore, we center the following discussion on these two overlooked dimensions: emotional and cognitive labor.

\subsubsection{From Designing for Visibility to Cultivating Child-Centered Mutual Understanding} 
Homework tutoring in Chinese households, despite rapid modernization, remains deeply gendered. Consistent with prior family research~\cite{daminger2019cognitive}, we found that mothers took the core cognitive labor of planning, monitoring, and adapting, alongside the intensive emotional labor of sustaining patience, motivating, and absorbing breakdowns~\cite{hochschild2012second}. Fathers, conversely, tend to play a secondary, auxiliary role (e.g., \say{optional helpers}). Their contributions were often fragmented and domain-specific, such as solving a difficult problem or cultivating educational values. This division seldom arises from explicit negotiation but rather accretes through the routine accommodation and cultural norms, naturalizing a state of \say{default maternal responsibility}~\cite{offer2014costs}. Consequently, the most depleting aspects of this labor, the constant planning and emotional regulation in the family, remain largely invisible and undervalued.

Our analysis reveals that the division of labor in homework tutoring extends far beyond the mere allocation of time or tasks. It uncovers a profound inequality in the division of emotional and cognitive labor~\cite{grandey2015emotional}. In our study, mothers were not only instructors but also emotional regulators, tasked with sustaining patience, sparking motivation, and mediating conflicts. They manage their children's frustration, anxiety, and resistance while simultaneously regulating their own emotional responses, a burden especially pronounced in high-pressure educational contexts.

The core cognitive labor of homework tutoring is also primarily carried by mothers. They predominantly served as project managers, anticipating challenges, planning pedagogical strategies, monitoring progress, and adapting flexibly to their child's evolving needs. Notably, our data reveal that parenting as a \say{strong norm} magnifies this gender gap: the demand for high-investment, child-centered care forces mothers to bear greater time and mental load costs. Consequently, this dynamic further entrenches the asymmetry of emotional and cognitive labor~\cite{hays1996cultural}.

We argue that invisible work in parenting is different from that in the workplace. The field of HCI increasingly recognizes the importance of making invisible emotional and cognitive labor visible, not to quantify and commodify it, but to foster healthier and more equitable relationships, both in professional and domestic spheres. Prior HCI and CSCW research has extensively examined invisible labor in workplace settings, often emphasizing how such labor can be made visible, quantified, and assigned value~\cite{toxtli2021quantifying, rosenblat2016algorithmic, meluso2025invisible}. However, the invisible labor that emerges within families differs from that in professional contexts. In the absence of a single, stable standard for judging such work, it is difficult to evaluate or commodify it in terms of labor value~\cite{dileonardo1987female, offer2014costs}. Unlike formal employment, where stability is enforced through hierarchical power and unequal evaluation rights, the domestic sphere functions differently. In intimate relationships centered on \say{shared family well-being}, spouses operate on an equal footing. Without the structural authority to command one-way obedience, conflicts in values cannot be dictated; they must be bridged through egalitarian deliberation, compromise, and the establishment of common norms.

In intimate family settings, fairness is not easily specified by a third party, and there is no single, stable standard against which labor can be judged. Technologies in the home are often entangled with gendered roles and with the everyday practices of care, maintenance, and repair. Prior work cautions against turning domestic spaces into sites of surveillance or audit and instead urges designers to support practices of upkeep, relational care, and ongoing maintenance from feminist and maintenance studies perspectives~\cite{Chambers2020DomesticatingSmarterThanYouHome}. Our findings suggest that when mothers express a desire for their labor to be \say{seen}, this wish is not fulfilled by simply surfacing more data about who does what. Rather, it reflects a desire for understanding and a pretext for negotiation. While one parent (typically the mother) often articulates a desire for equality, the precise definition of \say{fairness} in intimate contexts remains elusive. Given this ambiguity, we argue that making labor visible is not a matter of metrics, but a relational challenge that resists clear specification.

While the division of labor is often fraught with challenges, we observe that parental coordination is fundamentally anchored in a shared commitment to the child’s best interests. This triadic dynamic (father-mother-child), centering on the child’s well-being, elucidates why conflicts often evolve into processes of negotiation and mutual adaptation. Given that parents are united by a shared commitment to the child’s best interests, we propose using the child’s feedback to trigger the recognition of invisible labor, thereby facilitating the necessary renegotiation of parenting roles.

\subsubsection{The Struggle for Triadic Balance: Dynamic Cycles of Breakdown, Repair, and Renegotiation}
\begin{figure*}[t]
    \centering
    \includegraphics[width=0.8\linewidth]{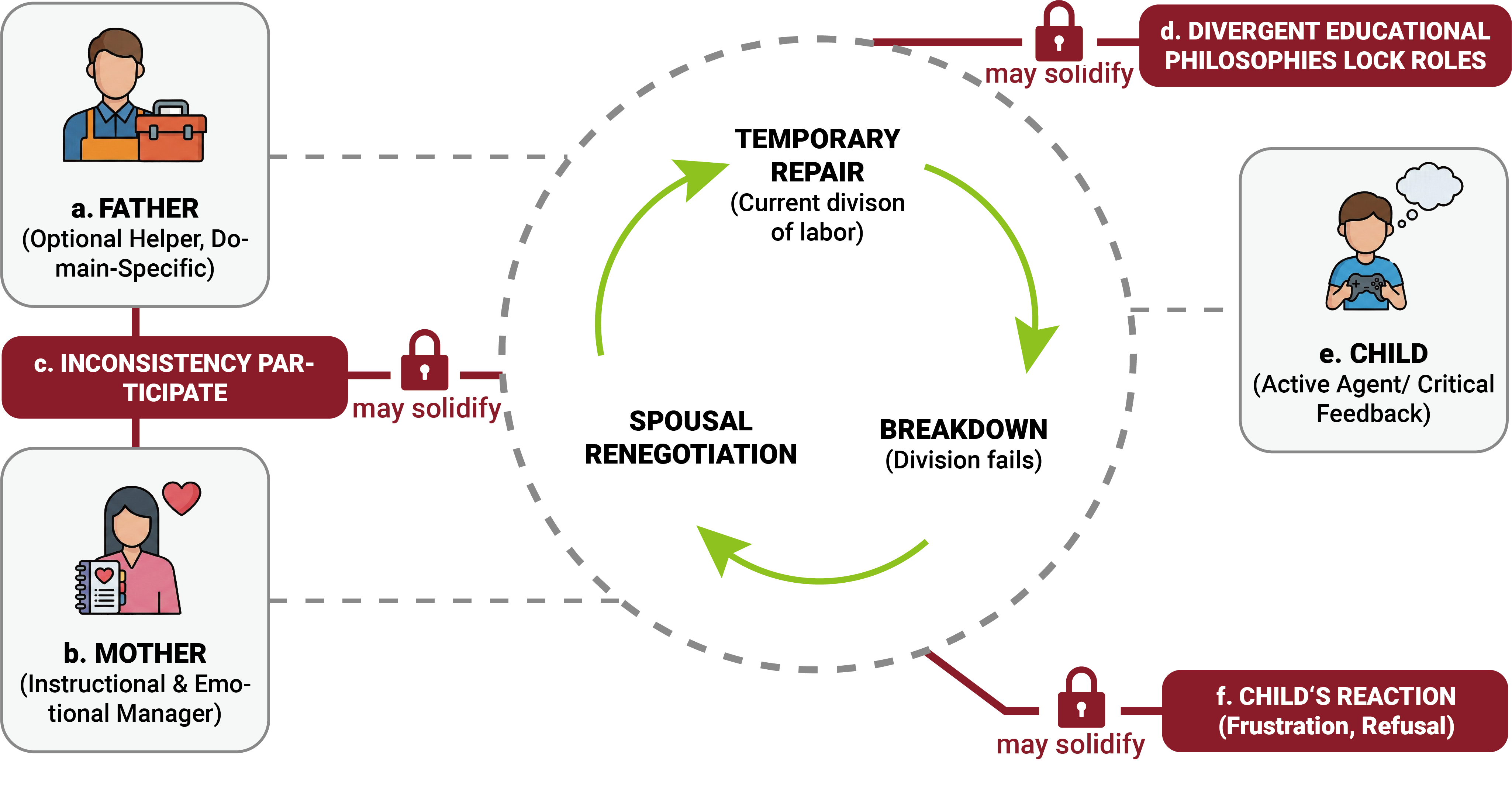}
    \caption{The model of Collaborative Breakdown and Repair in Homework Tutoring. \textbf{(a, b) Parental Roles}: The distinct roles of Father and Mother derive from structural constraints and gender norms (\textbf{Sec. 5.1.1, 5.1.2}). \textbf{(c, d) Locking Mechanisms}: \textit{Inconsistent Participation} and \textit{Divergent Educational Philosophies} act as barriers that solidify rigid labor divisions (\textbf{Sec. 5.1.2}). \textbf{(e) Child}: Acts as an active agent rather than a passive recipient (\textbf{Sec. 5.1.3}). \textbf{(f) Child's Reaction}: Frustration or refusal serves as the feedback loop triggering the breakdown cycle (\textbf{Sec. 5.1.3}).}
    \Description{A conceptual model of collaborative breakdown and repair in homework tutoring, showing parental roles, locking mechanisms, the child as an active agent, and the child’s reaction triggering renegotiation.}
    \label{fig:1}
\end{figure*}

Homework tutoring is fundamentally a triadic negotiation rather than a mere dyadic agreement between spouses. We found that children are not passive recipients of support but active agents who directly influence the allocation of parental responsibilities. Children’s emotional reactions, ranging from frustration to refusal, act as a critical feedback loop that tests the viability of the parents' arrangement. When a child resists a specific parent's approach, it triggers a breakdown in the current division of labor, necessitating immediate repair and renegotiation between spouses. Consequently, the child constitutes an integral part of the decision-making loop, driving the shift of responsibilities from one parent to the other.

As shown in Figure~\ref{fig:1}, the division of labor is not a static arrangement but a dynamic, recursive process grounded in our empirical findings. The cycle typically begins with a phase of Temporary Repair, reflecting unstable arrangements driven by the distinct roles of the Father (Fig. 1a) and Mother (Fig. 1b), as detailed in Section \ref{sec:5.1.1}. However, this stability is often fragile. As shown in Section \ref{sec:5.1.2}, conflicts arising from Inconsistent Participation (Fig. 1c) and Divergent Educational Philosophies (Fig. 1d) often act as a \say{lock}, preventing fluid role-switching and precipitating a Breakdown. Crucially, our model incorporates the Child (Fig. 1e) as an active agent; their specific Reaction (Fig. 1f), such as frustration or refusal detailed in Section \ref{sec:5.1.3}, often serves as the catalyst that forces parents into Spousal Renegotiation.

This reality complicates linear models like the \say{second shift}~\cite{dileonardo1987female}, where one caregiver, typically the mother, repeatedly absorbs disruptions and performs the relational repairs necessary to restore household equilibrium. Rather than viewing family labor solely as a predetermined burden shouldered by one member, our findings suggest that it acts as a dynamic cycle involving the entire family, including children. Within this framework, \say{breakdown} and \say{negotiation} are normalized as integral parts of family life. Ideally, this cycle allows for flexibility; however, in practice, it often exposes the fragility of shared parenting.

Maintaining this dynamic cycle is precarious. Our study reveals how situational crises, such as a child's resistance or a partner's inconsistency, transform \say{temporary fixes} into normalized obligations. This dynamic reflects mechanisms identified in prior CSCW research, whereby cognitive and emotional responsibilities become implicitly entrenched with the primary caregiver over time~\cite{daminger2019cognitive, hochschild2012second}. Our data adds a layer of complexity: divergent educational philosophies serve as the catalyst that locks these roles in place. The friction between mothers' focus on short-term performance and fathers' emphasis on long-term cultivation not only creates conflict but actively prevents effective task redistribution. As a result, what begins as a dynamic negotiation often solidifies into a static asymmetry, making the labor increasingly difficult to redistribute.

We therefore contend that sustaining a dynamic balance within the father-mother-child triad is inherently precarious, as these fluid interactions are prone to solidify into inequitable routines under the weight of daily friction. Consequently, rather than simply aiming to impose \say{clearer} roles, future interventions should focus on surfacing this invisible labor to facilitate value alignment better and support more effective renegotiation between parents.

\subsection{A Labor-Lens Design Agenda for AI in the Family}

We articulate an agenda for AI in the family that applies a labor lens. This approach challenges the tendency to design systems that judge \say{fairness}, as such framings often reduce care to visible, quantifiable tasks and marginalize less tangible forms of work~\cite{hochschild2012second,daminger2019cognitive}. Instead, we posit that the goal of AI technology should be to sustain the family system as a whole. We argue that the child’s feedback serves as the practical anchor for negotiation, positioning coparenting not as a dyadic struggle, but as a responsive triadic dynamic among father, mother, and child~\cite{feinberg2003coparenting}. Finally, we build on prior HCI work that demonstrates AI’s capacity to track, log, and summarize behavioral trajectories~\cite{luckin2016intelligence,holmes2019ai} by redirecting them from surveillance toward relational maintenance through collaborative annotation, narrative timelines, and joint reflection.

\subsubsection{Collaborative Annotation without Quantification.} 

Parents repeatedly asked for lightweight traces that they could revisit later. For instance, a mother (P8) described session logging that helps them review \say{where it flowed} and \say{where it stuck}, and capture the \say{hardest moment} or \say{most rewarding} points. This reflects a recurring need in our findings: parents are not seeking real-time advice as much as concrete anchors that can support later discussion and adjustment. The collaborative annotation involved three steps with appropriate AI involvement: capturing the environment, annotating information, and aggregating it into essential events to support both parents in understanding their context, coping style, and facilitating collaboration and communication~\cite{GaoEtAl2022SeatingExperience}. We frame collaborative annotations as a design space rather than a fixed interface prescription. The core idea is to provide caregivers with lightweight ways to mark where they feel \say{stuck} and where a \say{repair} seems to occur, so that these moments can be revisited when parents renegotiate responsibilities.

First, AI should help families capture small but critical moments of collaboration as they unfold, so that there is something concrete to communicate or care in the future. Their primary purpose is to provide anchors for later conversations about patterns of collaboration. This is particularly important when parents want to reflect on what happened without relying solely on memory or subjective impressions. Currently, AI has acquired capabilities such as microexpression~\cite{wang2025facebenchmultiviewmultilevelfacial}, emotion detection,~\cite{sasu2025ace,ahmadpour2025affective} emotion analysis~\cite{chen2024emotion,tak2024gpt4emulatesaveragehumanemotional,wu2025multimodalemotionrecognitionconversations}, environmental perception~\cite{qin2024mp5multimodalopenendedembodied,GaoEtAl2020nGage}, and even complex semantic understanding~\cite{lian2025affectgptnewdatasetmodel}. These capabilities can be expanded to proactively capture multimodal information in the family context, determining when to capture, what is important to capture, and how to capture it, according to specific family communication scenarios. Besides, we advocate regulating technological applications based on the core principle of \say{data minimization}, and also look forward to more in-depth discussions and consensus-building around this topic in the future HCI field to mitigate data concerns~\cite{Leschanowsky2025privacyQ&Adataset} in the family context~\cite{abdi2021privacy}.

Second, AI should enhance human-AI collaborative annotation. Research indicates that AI can effectively perform keyword annotation~\cite{gamage2025label}, understand context~\cite{gandhi2024humanlikeaffectivecognitionfoundation}, and integrate information. After capturing data in family scenarios, AI will generate initial labels covering basic aspects. Parents will then verify these labels, adding details about their family's unique context and their children's individual characteristics, allowing the AI to refine the annotations iteratively. This approach aligns with our findings that interpretations of family situations are highly situated and cannot be reliably inferred without parental input. In high-risk situations, such as child safety and ethical concerns~\cite{landesman2025children}, parental input and evaluation are crucial, and the limits of AI must be acknowledged. This collaborative approach mirrors successful human-AI partnerships in professional data annotation~\cite{gamage2025label} and strikes a balance between efficiency and safety in family contexts.


Third, the AI can support event aggregation to visualize the daily coparenting experience with potential causes of the diverse labor. By clustering breakdowns and repairs in the homework routine~\cite{Galbraith2024DialogueRepair,Li2024breakdowns,Alghamdi2024SystemUserStrategies}, it can reveal the types of work each parent typically takes on~\cite{gao2025homework,li2025learningtogetheraimediatedsupport, ho2024collaboration}. Here, the goal is not to enforce a notion of \say{fair distribution}, but to make contributions legible for coordination; as P8 noted, seeing \say{all the effort} can help parents identify \say{which parts could be shared}. Thus, AI should not only help parents engage more deeply in their children’s education and monitor related activities~\cite{garg2020conversational} but also identify unsustainable patterns in labor division through continuous data tracking. Instead of enforcing a standard for \say{fair distribution}, it offers a data-driven foundation for negotiating family responsibilities, particularly by making overlooked emotional and cognitive labor visible. Future research could explore human-AI collaborative models to better capture unique family interactions and labor characteristics.



\subsubsection{Child-Centered Narrative Timelines.} 

As a second design direction, we focus on AI constructing a child-centered narrative timeline rooted in the family’s overall goals, primarily manifested through the deep integration of children’s feedback into the entire design process. This direction is grounded in our findings that children’s reactions can repeatedly disrupt existing tutoring arrangements and prompt parents to reconsider how support should be provided. For example, a father (P13) described how his child’s persistent dissatisfaction made it clear that the current arrangement was not working and needed to be changed. Existing studies have confirmed that inferring a complete narrative from fragmented clues can effectively reduce users' cognitive load~\cite{wang2025cluecartsupportinggamestory}. One-size-fits-all and ill-considered parental control tools may fail to meet the diverse developmental needs of children across age groups~\cite{dumaru2025approriate}, and a child-centered design is better at identifying children’s needs and promoting their education.

First, we believe AI should effectively organize the event timelines, leveraging its strengths in collation and personalized insights~\cite{wang2025cluecartsupportinggamestory}. Existing research primarily focuses on single parents~\cite{jo2020understanding,freeman2023tracking,loerakker2025selftracking}, leaving out a comprehensive family perspective. We found that factors such as economic constraints and time availability significantly influence parental roles. For instance, a mother (P14) explained that frequent work travel made it practically impossible for her partner to participate consistently in tutoring, while another mother (P8) rationalized role allocation in terms of time and economic trade-offs. AI should integrate various information sources to highlight key contradictions in family collaboration, such as practical challenges (e.g., time conflicts), differing parenting values, and gender-based labor divisions (e.g., mothers facing more invisible tasks). At the same time, children’s moment-to-moment comprehension can still force adjustments within these constraints. As a father (P4) noted, when a child could not understand a particular explanation, parents would switch who would take over tutoring. Instead of merely identifying these issues, AI should provide objective explanations based on family collaboration theories and gender equality.
This approach can help families align their goals, adjust arrangements based on the timeline, and reduce emotional and cognitive labor, leading to more effective collaboration.

Second, AI should distinguish between children's actual needs and parents' perceived needs. Discrepancies often exist between what children truly need and how parents perceive those needs, especially among young children who may struggle to express themselves. Our findings suggest that children’s dissatisfaction often accumulates across sessions, rather than appearing as a single, decisive moment. For example, a father (P13) described how repeated misunderstandings during homework sessions gradually revealed deeper issues with how support was being provided. Thus, AI should capture and present children's needs objectively, rather than relying on parental assumptions. Existing studies illustrate that child-centered design fosters children's well-being and promotes collaboration among families~\cite{landesman2025children,garg2020conversational,Kim_2025}. AI can enhance this process by utilizing multimodal data fusion technologies, such as speech emotion recognition~\cite{Barhoumi2024RealTimeSER,Chowdhury2025SERLightweight}, behavioral trajectory analysis~\cite{Zeng2025PunishmentSensitivity,Yu2025InteractionDynamics}, and feedback recording~\cite{Han2025IntiVisor,Dodds2024ShowAndTell}. Research indicates that AI can help parents better understand their children~\cite{mei2025paceesupportingchildrenspersonal}, and there is potential for collaborative parenting between AI and parents~\cite{ho2024collaboration,garg2020conversational}. Additionally, AI should present feedback neutrally, avoiding binary \say{correct or incorrect} judgments. Instead, it should visualize the long-term impacts of addressing both \say{ought-to needs } and \say{is needs}, helping families make informed decisions.

\subsubsection{Joint Reflection and Co-Created Norms.} 
Event annotations and narrative timelines give families a record they can look at together, but records on their own do not repair relationships. Our findings suggest that parents want such records to serve as concrete anchors for discussion and renegotiation, rather than as passive archives. For example, a mother (P8) emphasized that seeing \say{all the effort} could help partners recognize contributions and identify \say{which parts could be shared}, and she framed AI as a \say{bridge in between} that supports discussion rather than judgment. We expect AI to guide parents in self-reflection, thereby fostering deeper mutual understanding~\cite{Stein2025MutualUnderstanding}. To achieve this goal, the design should focus on scaffolding dialogue and shared sense-making. Prior work has focused on mutual understanding for group collaboration~\cite{Schmutz2024AITeaming,He2024AIFutureCollaborativeWork}, such as facilitating knowledge sharing and perspective alignment in cross-functional project teams~\cite{Edwards2025HumanAICollaboration,Joshi2025ReviewAIManagement}. In family settings, AI has the potential to guide parents and offer targeted suggestions to foster mutual understanding.


We therefore argue that AI in the family should present objective information systematically and support parents in engaging in self-reflection~\cite{Bindayel2025AISelfReflection}. Existing research has confirmed that through functions such as behavioral data visualization and cognitive bias prompts, AI can act like a \say{mirror} to enhance individuals’ self-insight and cognitive awareness~\cite{Bindayel2025AISelfReflection}, which can be transferred to family contexts. This aligns with our findings that parents preferred facilitative, neutral framing for suggestions, so that feedback can be received as support for conversation rather than as a verdict. Meanwhile, AI possesses the potential to fulfill multiple roles, such as educator, inspirer, questioner, thinker, and recorder~\cite{hao2025mappingstudentaiinteractiondynamics}, which means it can transcend the attribute of a single tool and be deeply embedded in the complex family context, characterized by emotional bonds, power dynamics, and long-term interactions. We anticipate that AI in the family can flexibly switch roles in real time to achieve functional complementarity. For instance, through summary-based, question-based, heuristic, or hybrid interactive methods, it can subtly guide parents to independently reflect first about their behaviors and performance in mutual communication without undermining the naturalness of family interactions.

The reflection process should be carefully designed and implemented. Although the perspectives of the other party provided by AI are of reference value, they may also be transformed into evidence for disputes, ultimately leading to relationship breakdown rather than promoting cooperation. This concern resonates with our findings that parents explicitly wanted AI to reduce defensiveness and avoid escalating conflict during negotiation (e.g., a mother (P8) contrasted \say{you should} with \say{here’s what the data suggests}). Balancing the reflection design to support, yet not break, the family bond warrants further exploration.

\section{Limitations and Future Work} 
This study has several limitations that highlight opportunities for future research. A notable limitation of our study is the sociodemographic homogeneity of our participant sample. Specifically, our interviews focused on 18 parents of young children (grades 1–3) from primarily urban Chinese households. While this provided deep cultural insights, it limits the generalizability of our findings to more diverse family structures, such as single-parent or multigenerational households, where the division of labor and \say{invisible} care work may manifest differently~\cite{star1999invisible,daminger2019cognitive}. Future work should aim to collaborate with a broader range of communities to examine how factors such as income level and varying cultural backgrounds influence how parents negotiate the \say{livability} of their arrangements and achieve equitable divisions of labor.

Second, we used a single qualitative method: retrospective semi-structured interviews. While effective for understanding parents' perceptions, this approach risks missing the tacit, habitual nature of cognitive and emotional labor~\cite{daminger2019cognitive}, which often occurs in fleeting moments difficult to articulate fully. Furthermore, without in-situ observations, we could not capture the real-time dynamics of the triadic father-mother-child system~\cite{feinberg2003coparenting}, particularly the child’s immediate reactions that often trigger parental renegotiation. This leaves a potential gap between reported narratives and actual situated practices. Future research should employ mixed-method approaches, such as diary studies, longitudinal deployments, and direct involvement of children, to capture these granular dynamics as they evolve under actual academic pressures.

Furthermore, while we outlined design directions for AI as \say{relational support}, such as narrative timelines and event-based collaborative annotations, we have not yet developed or deployed these systems. Questions regarding technical feasibility, user burden, and potential unintended consequences (e.g., whether visualization fuels comparison rather than support~\cite{BoydAndalibi2023, roemmich2023emotion}) remain unexplored. Future work can focus on participatory co-design and iterative prototyping to test these concepts in the wild.

While we positioned AI to shift focus from task optimization to relationship maintenance, we acknowledge that the ethical implications of introducing data-driven technologies into the intimate space of the home require deeper investigation. Future research must rigorously examine privacy, data ownership, and the potential for surveillance to ensure that AI-generated insights truly scaffold shared understanding and family-level well-being without compromising the trust between family members.

\section{Conclusion}
This paper examines homework tutoring through a labor lens, showing that family collaboration is shaped less by visible tutoring tasks than by the often-unseen work of planning, tracking progress, and emotional repair, which can leave parents exhausted. These dynamics unfold within a triadic father-mother-child dynamic, in which the child’s feedback serves as the practical anchor for renegotiating roles. Therefore, we argue that addressing the imbalance in homework tutoring is not a problem of metrics or of defining \say{clearer} roles. Instead, it is a relational challenge that resists fixed specification and cannot be solved through performance optimization or judgments of objective fairness. AI in the family should therefore shift from surveillance and scorekeeping toward functioning as relational support of the family as a whole. We propose redirecting AI capabilities through three design directions: event-based collaborative annotation, narrative timelines, and joint reflection. By helping families surface hidden dynamics, align their values, and sustain context-sensitive negotiations, this work extends feminist and care-oriented HCI into domestic learning, positioning AI not as a referee but as a partner in sustaining equitable and durable caregiving arrangements.

\balance

\begin{acks}
This work is supported by the Natural Science Foundation of China (Grant No. 62302252).
\end{acks}
\bibliographystyle{ACM-Reference-Format}
\bibliography{sample-base}

\appendix


\end{document}